\documentclass[aps,prl,twocolumn,showpacs,superscriptaddress,groupedaddress]{revtex4}  
\usepackage{graphicx}  
\usepackage{dcolumn}   
\usepackage{amsmath,bm,amssymb,latexsym,galois,amsthm}   
\begin{document}

\widetext


\title{Monogamy Relation in No-disturbance Theories}
\author{Zhih-Ahn Jia\footnote{Email address: cajia@mail.ustc.edu.cn}}\affiliation{Key Laboratory of Quantum Information, Chinese Academy of Sciences, School of Physics, University of Science and Technology of China, Hefei, Anhui, 230026, P. R. China}
\affiliation{Synergetic Innovation Center of Quantum Information and Quantum Physics, University of Science and Technology of China, Hefei, Anhui, 230026, P. R. China}
\author{Yu-Chun Wu}\affiliation{Key Laboratory of Quantum Information, Chinese Academy of Sciences, School of Physics, University of Science and Technology of China, Hefei, Anhui, 230026, P. R. China}
\affiliation{Synergetic Innovation Center of Quantum Information and Quantum Physics, University of Science and Technology of China, Hefei, Anhui, 230026, P. R. China}
\author{Guang-Can Guo}\affiliation{Key Laboratory of Quantum Information, Chinese Academy of Sciences, School of Physics, University of Science and Technology of China, Hefei, Anhui, 230026, P. R. China}
\affiliation{Synergetic Innovation Center of Quantum Information and Quantum Physics, University of Science and Technology of China, Hefei, Anhui, 230026, P. R. China}

\date{\today}

\begin{abstract}
The monogamy is a fundamental property of Bell nonlocality and contextuality. In this article, we studied the $n$-cycle noncontextual inequalities and generalized CHSH inequalities in detail and found the sufficient conditions for those inequalities to be hold. According to those conditions, we provide several kind of tradeoff relations: monogamy of generalized Bell inequalities in non-signaling framework, monogamy of cycle type noncontextual inequalities and  monogamy between Bell inequality and noncontextual inequality in general no-disturbance framework. At last, some generic tradeoff relations of generalized CHSH inequalities for $n$-party physical systems, which are beyond one-to-many scenario, are discussed.
\end{abstract}

\pacs{}
\maketitle

\section{introduction}
The pioneering works by Bell\cite{Bell} and Kochen and Specker\cite{KS} shed light on the fact that the point of locally realistic interpretation and noncontextual realistic interpretation of the classical physics does not apply to the quantum world. Equivalently, there are no local hidden variable(LHV) and noncontextual hidden variable(NCHV) models for quantum mechanics. The simplest theoretical proofs of the nonexistence are described  mathematically by the violation of Clauser-Horne-Shimony-Holt(CHSH) inequality\cite{CHSH} and the violation of Klyachko-Can-Binicio\v{g}lu-Shumoviski(KCBS) inequality\cite{KCBS} respectively. A lot of designed experiments, like most recent loophole-free one\cite{Bell-exp}, demonstrated that we need reconsider the viewpoint of traditional truth on the basis of the viewpoint of holistic reality in quantum mechanics.\\
\indent The violation of CHSH inequality requires at least two spatially separated two dimensional parties, Alice and Bob, the correlation between them can be described as a joint probability distribution $p(a_i,b_j|A_i,B_j)$, where $A_i$ and $B_j$ are measurements chose by Alice and Bob, $a_i$ and $b_j$ are their outcomes respectively. There are some constraints on this correlation by non-signaling principle which says that signal can not travel faster than light(i.e. spatially separated parties can not affect each other), mathematically it reads $p(a_i|A_i)=\sum_{b_j}p(a_i,b_j|A_i,B_j)=\sum_{b_k}p(a_i,b_k|A_i,B_k)$, in which $A_i$ is the measurement chose by Alice and $B_j(B_k)$ is the measurement chose by Bob. These constraints guarantee that the measurements of Alice are independent of the other party's choices. One of the most fundamental properties of nonlocal correlations is monogamy relation. Scarani and Gisin\cite{m-CHSH1} showed that, for any three-qubit state shared by Alice, Bob and Charlie, if Alice and Bob do CHSH type experiment and get a violated value, namely $\langle CHSH_{A,B}\rangle\geq 2$, then the experiment result between Alice and Charlie must satisfy CHSH inequality, namely $\langle CHSH_{A,C}\rangle\leq 2$, more concisely $\langle CHSH_{A,B}\rangle+\langle CHSH_{A,C}\rangle \leq 4$, this monogamy relation can be derived from non-signaling principle solely\cite{m-CHSH2}. The result can be generalized for spatially separated Alice and $n$ Bobs\cite{m-Bell}, each Bob choose $n$ measurements to do the same Bell-type experiment with Alice, when Alice use the same $n$ measurements in each experiment, then these experiments can not simultaneously get the violated values. Ramanathan and Horodecki\cite{Bell-monogamy} give a more general result without restricting the number of measurements of each parties, but it still require that all inequalities in monogamy relation are of the same type and all Alice's measurement are shared in each inequality. Note that all above monogamy relations are in one-to-many scenario, i.e., Alice do experiments with several Bobs.\\
\indent The violation of KCBS inequality requires a single physical system whose dimension is larger than 3 or above and a set of measurements $\{A_1,\cdots,A_5\}$, where $A_i$ and $A_{i+1}(A_{5+1}=A_1)$ are compatible. In this case, non-signaling principle can be naturally generalized as no-disturbance principle, it states that for any measurements $A$ and $A_i(i=1,\cdots,m)$ such that $A$ is compatible with each $A_i$, then whichever $A_i$ is choosed to measure with $A$ jointly does not affect the distribution of $A$, reflecting on correlation probability distributions as: $\forall i,k=1,\cdots,m,p(a|A)=\sum_{a_i}p(a,a_i|A,A_i)=\sum_{a_k}p(a,a_k|A,A_k)$. Quantum mechanics satisfies the no-disturbance principle, but there exist theories that satisfy the no-disturbance principle and violate noncontextual and Bell inequalities more than quantum mechanics\cite{PR-box,m-Bell-KCBS}. Like Bell inequality, there are also some monogamy relations between KCBS inequalities\cite{m-KCBS}, but this relation is conditional, it requires two KCBS measurement sets have a compatible part. More interestingly, it is proved that there is monogamy relation between CHSH inequality and KCBS inequality, if these experiments share two measures\cite{m-Bell-KCBS}. Recently, an experimental verification of this monogamy relation is presented\cite{experiment-Bell-KCBS}.\\
\indent In this paper, we try to find the reason why monogamy occurs, so we stress the relationship between the monogamy relation and the local bound $R_L$(resp. contextual bound $R_C$), quantum bound(Tsirelson bound) $R_Q$, and non-signaling bound $R_{NS}$(resp. no-disturbance bound $R_{ND}$, we will not distinguish $R_{NS}$ and $R_{ND}$ if there is no ambiguity), and sufficient conditions of inequalities to be hold is derived. According to those conditions, we studied the generalized CHSH inequalities\cite{Gene-CHSH} and $n$-cycle noncontextual inequalities\cite{n-cycle-cont} with a view towards their monogamy aspects in detail. Compared with the work of Pawlowski and Brukner\cite{m-Bell} and the work of Ramanathan and Horodecki\cite{Bell-monogamy}, whose results are restricted on the same type Bell inequality with each party choosing the same number of  measurements in a one-to-many scenario, and all of Alice's measurements must be shared in each run of experiment, our results indicate that: (1) Bell inequality in each run of experiment is not necessarily of the same type, actually, there exist some classes of Bell ineuqalities, monogamy relations can be established in the same class; (2) It is not necessary to share all measurements chose by Alice, more explicitly, sharing two measurements are enough to reveal the monogamy of Bell nonlocality; (3) There exist monogamy relations beyond one-to-many scenario, which are more intrinsic from the entanglement viewpoint. Our results suggest that monogamy relation exists in some special class of inequalities, which may be utilized  to classify Bell nonlocality and contextuality.

\section{Compatible graphs, joint probability distributions and monogamy relations}
\begin{figure}
\includegraphics[scale=2.15]{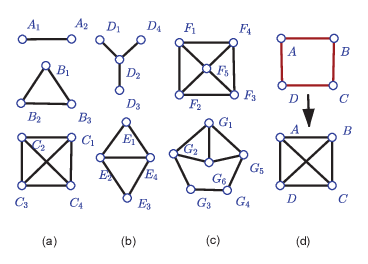}
\caption{\label{fig:graph}(color online). The black graph represent the compatible measurement graph, with the vertexes representing measurements and the edge representing the compatible relation. The red graph represent the experimental measurement graph, with the vertexes representing measurements involved in the experiment and the edge representing the joint measured relation. (a)complete compatible measurement graph, there naturally is a joint possibility distribution of all vertexes; (b)chordal compatible measurement graph, we can construct a joint possibility distribution such that all measurable marginal can be reproduced from this joint probability distribution; (c)inchordal compatible measurement graph; (d)the upper one is an experimental measurement graph, the graph below it is its corresponding compatible measurement graph.}
\end{figure}

For a set of measurements, we can draw a \emph{compatible measurement graph} $G(V,E)$, in which each measurement is represented by the a vertex in the vertices set $V$, the edge set $E\subseteq V^2$ contained all unordered pairs $(A_i,A_j)$ where $A_i$ and $A_j$ are compatible. Likewise, we can define the \emph{experimental measurement graph}, in which the vertex set consists of all measurements involved in one experiment, and for any two jointly measured measurements in the experiment, we connecting them by an edge, see Fig.~\ref{fig:graph}(a)-(c) for example. There is a compatible measurement graph corresponding to the experimental measurement graph, by connecting all compatible vertexes, in Fig.~\ref{fig:graph}(d) the upper one is an experimental measurement graph, in such an experiment, we measure the value of $\langle AB\rangle$, $\langle BC\rangle$, $\langle CD\rangle$ and $\langle DA\rangle$, but $A$, $B$, $C$ and $D$ are pairwise compatible, then we can draw its compatible graph as the lower graph in Fig.~\ref{fig:graph}(d). If $A_1,\cdots,A_n$ are a set of pairwise compatible measurements, i.e. its compatible measurement graph is complete, as in Fig.~\ref{fig:graph}(a), it is natural in physics to assumed that there exist a joint probability distribution $p(a_1,\cdots,a_n|A_1,\cdots,A_n)$ of all measurements, whose marginal are consistent with all partial measurements' probability distributions. For incomplete compatible measurements, people want to construct a joint probability distribution for the graph, such that all measurable marginal can be reproduced from this joint probability distribution. It was first found by Fine\cite{Fine} that in some case such a joint probability distribution can be constructed using some measurable marginal. To explain how to do this, see Fig.~\ref{fig:graph}(b), the joint probability distribution of upper graph can be constructed as $p(d_1,d_2,d_3,d_4)=p(d_1,d_2)p(d_3,d_2)p(d_4,d_2)/p^2(d_2)$, for the lower one, it is $p(e_1,e_2,e_3,e_4)=p(e_1,e_2.e_4)p(e_3,e_2,e_4)/p(e_2,e_4)$. Here, as well as the following section, we abbreviate the probability distribution $p(a|A)$ as $p(a)$ for convenience. Such a method is widely use in inequality monogamy problems and other probabilistic aspects of quantum theory\cite{m-KCBS,m-Bell-KCBS,test-cont}. Generally, it was proved that if the compatible measurement graph is a chordal graph, there is such a joint probability distribution\cite{m-KCBS}.\\
\indent \emph{Proposition~1}.
If the compatible measurement graph of a set of measurements is a chordal graph, then the graph admits a joint possibility distribution such that all measurable marginal can be reproduced from this joint probability distribution.\qed\\
\indent By chordal graph we mean a graph in which all cycles of four or more vertices have a chord, which is an edge that is not part of the cycle but connects two vertices of the cycle\cite{chordal graph}, equivalently, every induced cycle in the graph should have at most three vertices.\\
\indent From the viewpoint of quantum correlations, quantum theory is actually a non-signaling theory. In a specific run of the experiment, the correlations between each measurements are described by the joint probability distribution. For spatially separated observables, these correlations must satisfy the non-signaling conditions. Following the work of Acin, Gisin and Masanes\cite{m-CHSH2} let's recall the definition of non-signaling conditions. Consider spatially separated $n$-parties $\Gamma_n=\{S_1,\cdots,S_n\}$, each of them could measure their physical systems with different observables, then the correlations between these parties are described by a joint probability distribution $p(\Gamma_n)=\{p(s_1^{i_1}s_2^{i_2}\cdots s_n^{i_n}|S_1^{i_1}S_2^{i_2}\cdots S_n^{i_n})| S_k^{i_k}$ \emph{is the k-th party's measurement and $s_k^{i_k}$ is the corresponding outcome\}}. If the probability distribution of each subset of parties $\Gamma_m=\{S_{k_1},\cdots,S_{k_m}\}\subseteq \Gamma_n$ is the marginal probability distribution of some correlation $p(s_1^{i_1}s_2^{i_2}\cdots s_n^{i_n}|S_1^{i_1}S_2^{i_2}\cdots S_n^{i_n})$, or equivalently, the probability distribution $p(\Gamma_m)$ is independent of all $\Gamma_n-\Gamma_m$ parties, then we say that the correlations of these n-parties are non-signaling. It was proved that non-signaling conditions hold for all subset of $\Gamma_n$ if it holds for all subsets $\Gamma_{n-1}$\cite{non-signaling}. The importance of such a definition for yielding the monogamy relation of Bell inequalities will be illustrated latter in an example. \\
\indent No-disturbance conditions are  natural generalization of non-signaling conditions, these conditions concerns the compatible observables instead of only laying emphasis on spatially separated observables which are necessarily compatible. It states that, for any $n$ pairwise compatible observables $\Lambda_{n}=\{O_1,\cdots,O_n\}$, there is a joint probability distribution $p(\Lambda_n)=\{p(o_1,\cdots,o_n|O_1,\cdots,O_n)\}$, such that any distribution $p(\Lambda_m)$ of subset of observables $\Lambda_{m}=\{O_{k_1},\cdots,O_{k_m}\}\subseteq \Lambda_n$ is a marginal probability distribution of some $p(o_1,\cdots,o_n|O_1,\cdots,O_n)\in p(\Lambda_{n})$. No-disturbance conditions guarantee that the distribution $p(\Lambda_m)$ of measurement outcomes of some pairwise compatible observables are independent of the measurement of observables in $\Lambda_n-\Lambda_m$ which is compatible to all of the observables of $\Lambda_m$. Non-signaling conditions are a special kind of no-disturbance conditions, and quantum theory also satisfies the no-disturbance conditions, but there are theories that satisfy the no-disturbance conditions and violate non-contextuality inequalities more than quantum theory\cite{m-KCBS,PR-box}.\\
\indent To stress the importance of such a definition in Bell type monogamy problems, let us consider the monogamy relation between CHSH inequalities. For $CHSH_{AB}=A_1B_1+B_1A_2+A_2B_2-B_2A_1\leq 2$ and $CHSH_{AC}=A_1C_1+C_1A_2+A_2C_2-C_2A_1\leq 2$, the monogamy relation is
$CHSH_{AB}+CHSH_{AC}\leq 4$, if Alice and Bob share a Popescu-Rohrlich box(PR-box)\cite{PR-box}, Alice and Charlie also share a PR-box, it seems to violate the monogamy relation, but as we will show, this will lead a violation of our definition of non-signaling conditions.\\
\indent First we denote the probability distribution of two dichotomic measurements as
\begin{eqnarray*}
p(a_ib_j|A_iB_j)=\left(
  \begin{array}{cc}
    p(11|A_iB_j) & p(1-1|A_iB_j) \\
    p(-11|A_iB_j)& p(-1-1|A_iB_j)\\
  \end{array}
\right),
\end{eqnarray*}
actually, $p(a_1b_1|A_1B_1)$, $p(a_2b_1|A_2B_1)$, $p(a_2b_2|A_2B_2)$, $p(a_1c_1|A_1C_1)$, $p(a_2c_1|A_2C_1)$, and $p(a_2c_2|A_2C_2)$ have the same distribution:
\begin{eqnarray*}
\left(
  \begin{array}{cc}
    1/2 & 0 \\
    0 & 1/2 \\
  \end{array}
\right),
\end{eqnarray*}
while the probability distributions of $p(a_1b_2|A_1B_2)$ and $p(a_1c_2|A_1C_2)$ are the same:
\begin{eqnarray*}
\left(
  \begin{array}{cc}
    0 & 1/2 \\
    1/2 & 0 \\
  \end{array}
\right).
\end{eqnarray*}
\indent Since Alice, Bob and Charlie are spatially separated, there must be a set of distributions $p(\{B_j,A_i,C_k\})=\{p(b_ja_ic_k|B_jA_iC_k)\}$ satisfying nonsignaling conditions, but as we will show,this can not be achieved.\\
\indent Like two-measurement case, we denote the probability distribution of three measurements as
\begin{gather}
\begin{flalign*}
\begin{split}
p(b_ja_ic_k|B_jA_iC_k)=~~~~~~~~~~~~~~~~~~~~~~~~~~~~~~~~~~~~~~~~~~~\\
\left(
  \begin{array}{cccc}
      p(111) & & & p(11-1)\\
     &p(-111) &p(-11-1)& \\
     &p(1-11) &p(1-1-1)& \\
     p(-1-11) & & &p(-1-1-1)\\
  \end{array}
\right).
\end{split}
\end{flalign*}
\end{gather}
 \indent By solving linear equations we know that $p(B_1A_1C_1)$, $p(B_1A_2C_1)$, $p(B_1A_2C_2)$, $p(B_2A_2C_1)$ and $p(B_2A_2C_2)$ have the same distribution, and the distributions of all $p(B_jA_iC_k)$ are
 \begin{eqnarray*}
\left(
  \begin{array}{cccc}
      1/2 & & & 0\\
     &0 &0& \\
     &0 &0& \\
     0 & & &1/2\\
  \end{array}
\right)_{\begin{subarray}{lcl}
           jik\\
           i+j\not=3\\
           i+k\not=3
         \end{subarray}}
\left(
  \begin{array}{cccc}
      0 & & & 0\\
     &1/2 &0& \\
     &0 &1/2& \\
     0 & & &0\\
  \end{array}
\right)_{211},
\end{eqnarray*}
\begin{eqnarray*}
\indent\left(
  \begin{array}{cccc}
      0 & & & 1/2\\
     &0 &0& \\
     &0 &0& \\
     1/2 & & &0\\
  \end{array}
\right)_{112}\indent\indent
 \left(
 \begin{array}{cccc}
      0 & & & 0\\
     &0 &1/2& \\
     &1/2 &0& \\
     0 & & &0\\
  \end{array}
\right)_{212}.
\end{eqnarray*}
\indent Using all these messages, we get marginal probability distribution like
\begin{eqnarray*}
\sum_{a_1}p(b_1,a_1,c_2|B_1,A_1,C_2)=
\left(
  \begin{array}{ccc}
    0 & 1/2 \\
    1/2 & 0 \\
  \end{array}
\right),
\end{eqnarray*}
\begin{eqnarray*}
\sum_{a_2}p(b_1,a_2,c_2|B_1,A_2,C_2)=
\left(
  \begin{array}{ccc}
    1/2 & 0 \\
    0 & 1/2 \\
  \end{array}
\right).
\end{eqnarray*}
This is obviously contradict with non-signaling condition. Note that this result is an example of monogamy between PR-box, i.e. two PR-boxes can not be simultaneously shared by Alice-Bob and Alice-Charlie. Monogamy relation of nonlocalities and contextualies have a strong relation with the monogamy of quantum probability boxes.\\
\indent We now explain how these definitions are related with monogamy of various inequalities. For two inequality-type nonlocality(or contextuality) testing experiments $I_i\leq R_{L_i}\leq R_{Q_i}\leq R_{NS_i}(i=1,2)$, where $R_{L_i}, R_{Q_i}$ and $R_{NS_i}$ are their local bounds, quantum bounds and non-signaling bounds respectively. We can draw the experimental measurement graphs of these two inequalities and their corresponding compatible graphs which are usually some isolated incomplete graphs(if the compatible graph is a complete graph, then $R_{L}= R_{Q}= R_{NS}$, it can not test anything). For these two experiments, the monogamy relation reads $I=I_1+I_1\leq R_L=R_Q=R_{L_1}+R_{L_2}$, i.e. the summed inequality's local bound and quantum bound are the same, both are equal to the summation of two initial local bound, then, only one of the inequality can get violated value, this is the monogamy relation. For our purpose, we only concern the case that two incomplete graphs are connected in some points, namely, two inequalities share some measurements(if not, two isolated graphs have no interaction, the monogamy relation $I=I_1+I_1\leq R_L=R_Q=R_{L_1}+R_{L_2}$ holds only if $R_{L_1}=R_{Q_1}$ and $R_{L_2}=R_{Q_2}$, the relation $R_L=R_Q$ between local bound and quantum bound of summed inequality holds trivially). If the overall compatible graph of two inequalities is a chordal graph, then there exist a joint probability distribution which can reproduce all measurable correlations involved in two experiment, then both of them can not get violated value, the result is trivial. It is natural to consider the subgraph of the overall compatible graph, we can recombined the inequalities as $I_1+I_2=I^{(1)}+I^{(2)}$ with local bound satisfying $R_{L^{(1)}}+R_{L^{(2)}}=R_{L_1}+R_{L_2}$, if both compatible graphs of $I^{(1)}$ and $I^{(2)}$ are chordal graph, then we can prove the monogamy relation. This is exactly what we will do in the following section.

\section{Main Results}
The experimental measurement graph of Genelized CHSH inequalities\cite{Gene-CHSH}, $\mathcal{B}(2m)=A_1B_1+B_1A_2+A_2B_2+\cdots+A_{m-1}B_m-B_mA_{1}\leq 2m-2$, is a $2m$-cycle. In KCBS-type non-contextuality testing scenario, the experimental measurement graph of $\mathcal{C}(5)=A_1A_2+A_2A_3+\cdots+A_5A_1\geq -3$ is also a cycle. These inequalities can be generalized as a unified $n$-cycle noncontextual inequality\cite{n-cycle-cont} with measurements $\{A_1,\cdots, A_n\}$:
\begin{eqnarray}
\langle\mathcal{C}(n)\rangle &=&\sum_{i=1}^{n}\gamma_i\langle A_iA_{i+1}\rangle, \nonumber\\
&\leq_{C}& n-2,\nonumber\\
&\leq_{Q}&\left\{
\begin{array}{lcl}
\frac{3n\cos(\frac{\pi}{n})-n}{1+\cos(\frac{\pi}{n})}, n\in 2\mathbb{N}+1\\
n \cos(\frac{\pi}{n}),n\in 2\mathbb{N}
\end{array}\right.,\nonumber\\
&\leq_{NS}&n,
\end{eqnarray}
where $A_{n+1}=A_1$, $\gamma_i=\pm1$ and the number of negative $\gamma_i$ is odd. Since we can substitute $A$ with $-A$, the number of negative $\gamma_i$ can been reduced to one.\\
\indent\emph{Theorem~1}.
For any $n$-cycle noncontextual(resp. 2m-generalized CHSH) experiment with measurements $\{A_1,\cdots,A_n\}$(resp. $\{A_1,\cdots,A_m,B_1,\cdots,B_m\}$), If there is a joint probability distribution for all involved measurements $p(a_1,\cdots,a_n|A_1,\cdots,A_n)$(resp. $p(a_1,\cdots,a_m,b_1,\cdots,b_m|A_1,\cdots,A_m,B_1,\cdots,B_m)$) such that all measurable marginal can be reproduced from this joint probability distribution, then the inequality holds: $\mathcal{C}(n)\leq n-2$(resp. $\mathcal{B}(2m)\leq 2m-2$).\\
\indent\emph{Proof}. Suppose $p(a_1,\cdots,a_n)$ is such a joint possibility distribution, then
\begin{eqnarray*}
\langle\mathcal{C}(n)\rangle&=&\sum_{a_1,a_2}\gamma_1a_1a_2p(a_1a_2)+\cdots+\sum_{a_n,a_1}\gamma_na_na_1p(a_na_1),\\
&=&\sum_{a_1,\cdots,a_n}(\gamma_1a_1a_2+\cdots+\gamma_na_na_1)p(a_1,\cdots,a_n),\\
&\leq & \Big|\sum_{a_1,\cdots,a_n}(\gamma_1a_1a_2+\cdots+\gamma_na_na_1)p(a_1,\cdots,a_n)\Big|,\\
&\leq & \sum_{a_1,\cdots,a_n}\big|(\gamma_1a_1a_2+\cdots+\gamma_na_na_1)\big| p(a_1,\cdots,a_n),\\
&\leq & \sum_{a_1,\cdots,a_n}\max\{|\gamma_1a_1a_2+\cdots+\gamma_na_na_1|\}\\
&\indent&\indent\indent\indent\times p(a_1,\cdots,a_n),\\
&\leq & \sum_{a_1,\cdots,a_n}(n-2)p(a_1,\cdots,a_n),\\
&=&n-2.
\end{eqnarray*}
\indent For generalized CHSH inequalities, the proof is just the same.\qed\\
\indent Since chordal compatible measurement graph have the joint possibility distributions meet the requirements, each inequality experimental measurement graph whose compatible measurement graph is a chordal graph can not get the violated values in no-disturbance(non-signaling) theories.\\

\begin{figure}
\includegraphics[scale=1.07]{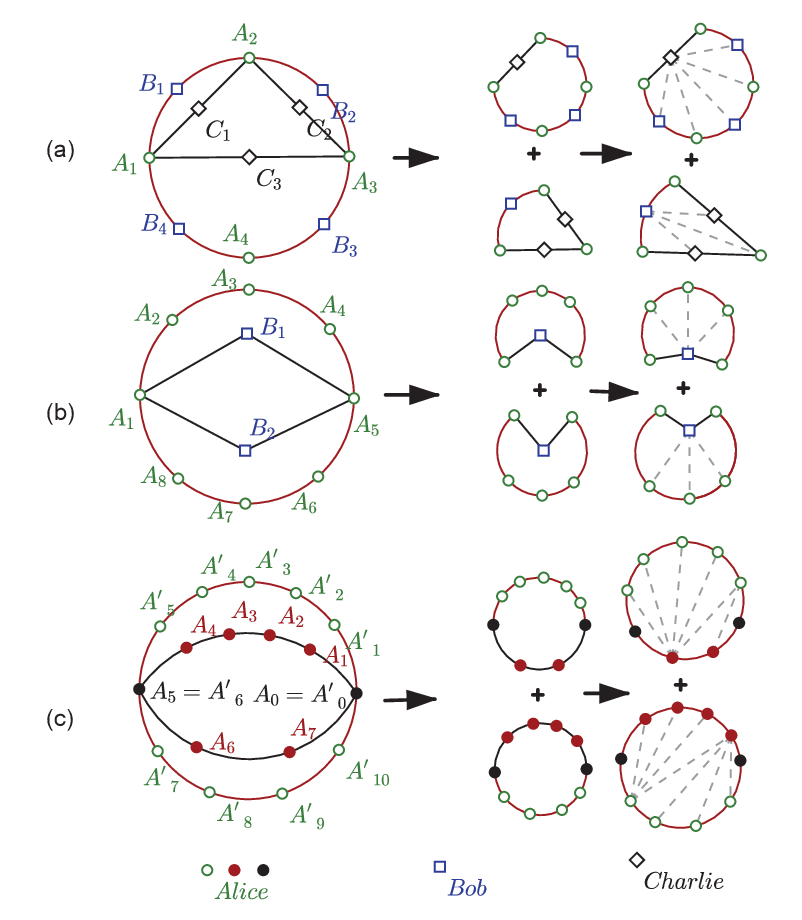}
\caption{\label{fig:proof} (color online). The sketch of proofs of the theorem 2, theorem 3 and theorem 4. On the left side of the figure, its the experimental measurement graph of each case, in the middle it's the graph of each recombined form of inequality, and on the right side it's compatible graph of each recombined inequality graph. (a)The sketch of proof of monogamy relation of $\mathcal{B}(2\times4)+\mathcal{B}(2\times3)\leq R_{L_8}+R_{L_6}=6+4$. (b)The sketch of proof of monogamy relation of $\mathcal{C}(8)+CHSH\leq R_C+R_L=6+2$. (c)The sketch of proof of $\mathcal{C}(8)+\mathcal{C}(11)\leq R_{C_8}+R_{C_{11}}=6+9$.}
\end{figure}

\indent\emph{Theorem~2}. Suppose Alice do Bell nonlocal experiments with $k$ Bob simultaneously in a generalized CHSH scenario, if these experiments share at least two measurements in Alice's part, then there is a monogamy relation between these $k$ generalized CHSH inequalities, mathematically, it reads
\begin{eqnarray}
\mathcal{B}_{1}(2m_1)+\cdots+\mathcal{B}_{k}(2m_k)&\leq& R_{L_1}+\cdots+R_{L_k},\nonumber\\
&=&\sum_{i=1}^k(2m_i)-2k,
\end{eqnarray}
in which $R_{L_i},i=1,\cdots,k$ is locality bound.\\
\indent\emph{Proof}. Let $\mathcal{B}_k(2m_k)=A_1^kB_1^k+B_1^kA_2^k+\cdots+A_{m_k}^kB_{m_k}^k-B_{m_k}^kA^k_{1}\leq R_{L_k}=2m_k-2$. Without loss of generality, suppose $A_1^i=A_1$ and $A_2^i=A_2$ for all $i=1,2,\cdots,k$. Let $\mathcal{B}^{(i)}=A_1B_1^{j}+B_1^{j}A_2^i+A_2^iB_2^i+B_2^i+\cdots+A^i_{m_i}B^i_{m_i}-B^i_{m_i}A_1^i$, in which $i\not= j$, $i$ and $j$ are chose such that the number of negative terms is exactly one. Since the local bound of generalized CHSH inequality only depends on the number of measurements, $m_i=m^{(i)}$, thus $R_{L_i}=R_{L^{(i)}}$. Since all $R_{L_i}$ and $R_{L^{(i)}}$ are positive, to violate the monogamy relation, at least one of recombined Bell inequality should be violated, say $\mathcal{B}^{(i)}>R_{L^{(i)}}$. However, the compatible graph of $\mathcal{B}^{(i)}$ is a chordal graph(for measurement $B^j$ is compatible with all other measurement in $\mathcal{B}^{(i)}$), there exists a joint possibility distribution such that all measurable marginal can be reproduced from it, by theorem 1, the inequality can not be violated, this is a contradiction. Thus, we have $\sum_{i=1}^k\mathcal{B}_i(2m_i)=\sum_{i=1}^k\mathcal{B}^{(i)}(2m^{(i)})\leq \sum_{i=1}^k R_{L^{(i)}}= \sum_{i=1}^k(2m^{(i)}-2)=\sum_{i=1}^k(2m_i-2)=\sum_{i=1}^k R_{L_i}$.
\qed\\
\indent To illustrate the method more explicitly, we give the proof for the case $\mathcal{B}(2\times4)+\mathcal{B}(2\times3)\leq R_{L_8}+R_{L_6}=6+4$, see Fig.~\ref{fig:proof}(a) for the sketch. Suppose that Alice and Bob do Bell measurements $\mathcal{B}(2\times4)=A_1B_1+B_1A_2+A_2B_2+B_2A_3+A_3B_3+B_3A_4+A_4B_4-B_4A_1\leq6$, while Alice and Charlie measure $\mathcal{B}(2\times3)=A_1C_1+C_1A_2+A_2C_2+C_2A_4+A_4C_3-C_3A_1\leq4$. Recombining each joint term in $\mathcal{B}(2\times4)+\mathcal{B}(2\times3)$, we get $\mathcal{B}^{(1)}(2\times 4)=A_1B_1+B_1A_2+A_2B_2+B_2A_3+A_3B_3+B_3A_4+A_4C_3-C_3A_1\leq6$, with $\mathcal{B}^{(2)}(2\times3)=A_1C_1+C_1A_2+A_2C_2+C_2A_4+A_4B_4-B_4A_1\leq4$. To violate inequality $\mathcal{B}^1+\mathcal{B}^2=\mathcal{B}^{(1)}+\mathcal{B}^{(2)}\leq 6+4$, either $\mathcal{B}^{(1)}\leq6$ or $\mathcal{B}^{(2)}\leq4$ should be violated, however,  since both of their corresponding compatible measurement graphs are chordal graphs, we can construct the joint probability distributions as:
\begin{gather*}
\begin{flalign}
\begin{split}
p(a_1c_1a_2b_2a_3b_3a_4b_4)=~~~~~~~~~~~~~~~~~~~~~~~~~~~~~~~~~~~~~~~~~~~~~\\
\frac{p(a_1c_1b_4)p(c_1b_4a_4)p(c_1a_4b_3)p(c_1b_3a_3)p(c_1a_3b_2)p(c_1b_2a_2)}{p(c_1b_4)p(c_1a_4)p(c_1b_3)p(c_1a_3)p(c_1b_2)},\nonumber\\
p(a_1b_1a_2c_2a_3c_3)=\frac{p(a_1b_1c_3)p(b_1c_3a_3)p(b_1a_3c_2)p(b_1c_2a_2)}{p(b_1c_3)p(b_1a_3)p(b_1c_2)}.\nonumber\
\end{split}
\end{flalign}
\end{gather*}
Every experimental joint measurement probability distribution can be reproduced from these two distributions, thus in non-signaling theory, they both can not get violated values, this is a contradiction. \\
\indent Note that for $k$ inequalities, we can only recombine them into $k$ inequalities with chordal compatible measurement graph, otherwise the upper bounds will change.\\
\indent This result show that Bell nonlocal monogamy may exist in different type Bell inequalities, it seems that this monogamy relation can be used to classify Bell nonlocality. \\
\indent\emph{Theorem~3}. For any contextual type experiment $\mathcal{C}(N)$ of Alice and CHSH type experiment $CHSH$ between spatially separated Alice and Bob, with two measurements $A_1$ and $A_2$ shared in both experiments, there is such a monogamy relation:
\begin{eqnarray}
\mathcal{C}(N)+CHSH\leq R_C+R_L=N-2+2=N.
\end{eqnarray}
\indent\emph{Proof}. This is the same sa proof of the above theorem, detailed process for $\mathcal{C}(8)+CHSH\leq R_C+R_L=8$ is sketched in Fig.~\ref{fig:proof}(b).
\qed\\
\indent \emph{Theorem~4}. For any $k$ contextual inequalities $\mathcal{C}(N_i)=A^{(i)}_1A^{(i)}_2+A^{(i)}_2A^{(i)}_3+\cdots+A^{(i)}_{N_i}A^{(i)}_1\leq R_{C_i},i=1,\cdots,k$, with two measurements sharing in each inequality and any pair of measurements(excluding the sharing two measurements) from different inequalities are compatible. Then there is a monogamy relation between these inequalities:
\begin{eqnarray}
\sum_{i=1}^k\mathcal{C}(N_i)\leq \sum_{i=1}^k R_{C_i}.
\end{eqnarray}
\indent\emph{Proof}. The proof is the same as the proof of theorem 2, detailed process for $\mathcal{C}(8)+\mathcal{C}(11)\leq R_{C_8}+R_{C_{11}}=6+9$ is sketched in Fig.~\ref{fig:proof}(c), as in the figure, any $A_i(i\neq 0,5)$ is compatiable with $A'_j(j\neq 0,6)$.
\qed\\
\indent\emph{Theorem~5}. For $N$ spatially separated parties $\{S_i\}_{i=1,\cdots,N}$ and $N$ generalized CHSH type Bell experiments $\mathcal{B}_{i,i+1}$(particularly $CHSH_{i,i+1}$) between $i$-th party and $(i+1)$-th party, with $\mathcal{B}_{i,i+1}$ and $\mathcal{B}_{i-1,i}$ having at least two measurements in common, we have:\\
\indent(1)Loop-type many-party monogamy relation:
\begin{eqnarray}
\mathcal{B}_{1,2}+\mathcal{B}_{2,3}+\cdots+\mathcal{B}_{N-1,N}+\mathcal{B}_{N,1}\leq \sum_{i=1}^N R_{L_{i,i+1}},
\end{eqnarray}
where $N+1=1$, each $R_{L_{i,i+1}}$ is the local bound of generalized CHSH inequality $\mathcal{B}_{i,i+1}$.\\
\indent(2)Chain-type many-party monogamy relation:
\begin{eqnarray}
CHSH_{1,2}+\cdots+CHSH_{2n,2n+1} &\leq&R_L,\nonumber\\
&=&R_Q\nonumber,\\
&=&R_{NS}=4n
\end{eqnarray}
where $R_L$, $R_Q$ and $R_{NS}$ are the local bound, quantum bound and non-signaling bound of sumed CHSH inequality respectively.\\
\indent\emph{Proof}. (1)We can equivalently consider the violation of $2(\mathcal{B}_{1,2}+\mathcal{B}_{2,3}+\cdots+\mathcal{B}_{N,1})\leq 2\sum_{i=1}^N R_{L_{i,i+1}}$, to get the violated value, at least one couple of $\mathcal{B}_{i-1,i}+\mathcal{B}_{i,i+1}\leq R_{L_{i-1}}+R_{i}$ should be violated. However, from theorem 2 we know $\mathcal{B}_{i-1,i}$ and $\mathcal{B}_{i,i+1}$ are monogamy for all $1=1,\cdots,N$, thus we arrive a contradiction. As an example, consider Alice, Bob and Charlie are doing CHSH experiments pairwise, with $CHSH_{AB}=A_1B_1+B_1A_2+A_2B_2-B_2A_1$, $CHSH_{BC}=B_1C_1+C_1B_2+B_2C_2-C_2B_1$ and $CHSH_{CA}=C_1A_1+A_1C_2+C_2A_2-A_2C_1$. To violate the inequality $(CHSH_{AB}+CHSH_{BC})+(CHSH_{BC}+CHSH_{CA})+(CHSH_{CA}+CHSH_{AB})\leq 12$, at least one $(CHSH_{i,i}+CHSH_{j,k})\leq 4(i,j,k=A,B,C)$ should be violated, this is impossible from the monogamy relation of CHSH inequalities.\\
\indent(2)Note that each two adjacent CHSH inequalities are monogamous, we can let one of it get the maximal value, say 2. By this way we see that the theorem is a direct consequence of the monogamy relation of CHSH inequalities. Note that the number of CHSH inequalities must be odd, otherwise, there is no monogamy relation. For example, sumed CHSH inequality $CHSH_{1,2}+\cdots+CHSH_{2n-1,2n}\leq_L 4n-2\leq_{Q}R_Q\leq_{NS}4n$ have different local bound and quantum bound.
\qed\\
\indent The above theorem is two kinds of tradeoff relations beyond one-to-many scenario, they are somewhat more intrinsic from the viewpoint of correlations of many-party physical systems. Qin \emph{et al.}\cite{3-loop} have a very similar result with our loop-type monogamy relation for three patries, expressed with sum of squares of CHSH operators' expectation values: $\langle CHSH\rangle_{AB}^2+\langle CHSH\rangle_{BC}^2+\langle CHSH\rangle_{CA}^2\leq 12$.\\
\section{conclusions}
In this article, we present a method which is suitable for a large set of monogamy problems, such as cycle type noncontextual inequalities and generalized CHSH inequalities. For a nonlocality testing experiment(resp. contextuality testing experiment), the existence of joint possibility distribution, with which all measurable marginal can be reproduced from it, prevents the violations of nonlocal inequality(resp. noncontextual inequality). Utilizing this method, we present several novel monogamy relations of one-to-many type: monogamy of generalized CHSH inequalities(not necessarily of the same type and only sharing two measurements); monogamy of $n$-cycle noncontextual inequalities(not necessarily of the same type and only sharing two measurements); and monogamy between CHSH inequalities and $n$-cycle noncontextual inequalities. Besides, we exhibit some tradeoff relations like loop-type and chain-type monogamies of generalized CHSH inequalities for $n$ parties beyond one-to-many scenario.\\
\indent Z.-A. J. acknowledges Bai-Chu Yu, Biao Yi and Sheng Tan for many beneficial discussions. This work is supported by the National Basic Research Program of China (Grant Nos. 2011CBA00200 and 2011CB921200), the Strategic Priority Research Program of the Chinese Academy of Sciences (Grant Nos. XDB01030100,XDB01030300), the National Natural Science Foundation of China (Nos. 11275182, 61435011).

\end{document}